\documentclass[conference]{IEEEtran}

\IEEEoverridecommandlockouts
\usepackage{cite}
\usepackage{amsmath,amssymb,amsfonts}
\usepackage{algorithmic}
\usepackage{graphicx}
\usepackage{textcomp}
\usepackage{xcolor}
\def\BibTeX{{\rm B\kern-.05em{\sc i\kern-.025em b}\kern-.08em
    T\kern-.1667em\lower.7ex\hbox{E}\kern-.125emX}}

\usepackage{tcolorbox}
\usepackage{adjustbox}
\usepackage{booktabs}
\usepackage{multirow}
\usepackage{tabularx}
\usepackage[colorlinks=true, allcolors=black]{hyperref}

\makeatletter
\def\ps@IEEEtitlepagestyle{
\let\@oddhead\@empty
\let\@evenhead\@empty
\def\@oddfoot{\footnotesize 979-8-3315-9147-2/25/\$31.00 ©2025 IEEE\hfill}%
\let\@evenfoot\@empty
}
\makeatother

\begin{document}

\title{What About Our Bug? A Study on the Responsiveness of NPM Package Maintainers }

\author{
    \IEEEauthorblockN{Mohammadreza Saeidi}
    \IEEEauthorblockA{
        \textit{University of British Columbia} \\
        BC, Canada \\
        mohammadreza.saeidi@ubc.ca
    }
    
    \and

    \IEEEauthorblockN{Ethan Thoma}
    \IEEEauthorblockA{
        \textit{University of British Columbia} \\
        BC, Canada \\
        ethan.thoma@ubc.ca
    }

    \and

    \IEEEauthorblockN{Raula Gaikovina Kula}
    \IEEEauthorblockA{
        \textit{University of Osaka} \\
        Osaka, Japan \\
        raula-k@ist.osaka-u.ac.jp 
    }

    \and

    \IEEEauthorblockN{Gema Rodríguez-Pérez}
    \IEEEauthorblockA{
        \textit{University of British Columbia} \\
        BC, Canada \\
        gema.rodriguezperez@ubc.ca
    }
}
\maketitle

\begin{abstract}
Background: Widespread use of third-party libraries makes ecosystems like Node Package Manager (npm) critical to modern software development. 
However, this interconnected chain of dependencies also creates challenges: bugs in one library can propagate downstream, potentially impacting many other libraries that rely on it. We hypothesize that maintainers may not always decide to fix a bug, especially if the maintainer decides it falls out of their responsibility within the chain of dependencies.
Aims: To confirm this hypothesis, we investigate the responsiveness of 30,340 bug reports across 500 of the most depended-upon npm packages.
Method: We adopt a mixed-method approach to mine repository issue data and perform qualitative open coding to analyze reasons behind unaddressed bug reports. 
Results: Our findings show that maintainers are generally responsive, with a median project-level
responsiveness of 70\% (IQR: 55\%–89\%), reflecting their commitment to support downstream developers. 
Conclusions: We present a taxonomy of the reasons some bugs remain unresolved.  The taxonomy included contribution practices, dependency constraints, and library-specific standards as reasons for not being responsive. Understanding maintainer behavior can inform practices that promote a more robust and responsive open-source ecosystem that benefit the entire community.
\end{abstract}

\section{Introduction} 
\label{sections:introduction}
Third-party packages play a crucial role in modern software development by reducing implementation effort and providing reusable functionalities that would otherwise be time-consuming to develop from scratch~\cite{gaikovina2023promises}. Package managers simplify the integration of these packages into projects. In the JavaScript ecosystem, the Node Package Manager (npm) serves as a vital infrastructure component~\cite{hafner2021node}. As of 2019, 72\% of npm packages relied on at least one third-party package~\cite{vaidya2019security}, and by 2023, the most depended-upon package, body-parser, had over 850,000 direct dependents~\cite{Wattanakriengkrai2023}, highlighting the deeply interconnected nature of the npm ecosystem.

However, this interconnectedness also introduces challenges. Bugs in widely used packages can cascade through the dependency chain, causing crashes, vulnerabilities, or performance issues for numerous downstream systems~\cite{derr2017keep, huang2022characterizing}. Addressing such issues often requires effective collaboration between \textbf{upstream developers (who maintain the packages) and downstream developers (who consume them)}.

Despite the importance of this collaboration, we still lack a clear understanding of how—and to what extent—upstream developers engage in fixing bugs. Do maintainers actively attempt to resolve reported issues? Why do some bug reports go unaddressed? Who is ultimately responsible for fixing the bug? These questions are especially critical for highly depended-upon npm packages, where unresolved bugs can compromise the reliability of the broader ecosystem~\cite{haq2025ripple, przymus2025out}.

Evaluating upstream engagement is not straightforward. While upstream developers are ideally expected to respond to GitHub issues submitted by users, this expectation is complicated by the fact that issues often include feature requests or performance suggestions~\cite{herzig2013s, heppler2016cares}, not just bug reports. One key challenge is determining bug ownership, whether the upstream or downstream developer is ultimately responsible for resolving an issue. Not all reported bugs are actual defects; some result from misunderstandings, environment-specific factors, or unrealistic expectations~\cite{herzig2013s, rodriguez2016bugtracking, panichella2021won}. Another challenge is assessing bug resolution, as the closure of an issue does not necessarily indicate that the problem has been fixed~\cite{wang2019my, khatoonabadi2023wasted, li2021you}. In some cases, upstream developers may close issues without resolving the underlying problem, because downstream developers fail to address the upstream feedback, or upstream developers do not follow up on the issue's progress. Hence, relying only on issue status can lead to misleading conclusions about maintainer behavior.

In this study, we consider both bug ownership and actual resolution status as key dimensions for \textbf{evaluating the level of responsiveness of npm package maintainers to bug reports (what)}. We define responsiveness as the extent to which upstream developers resolve bug reports for which they are responsible. To capture this, we propose a \textbf{framework and automated approaches (how)} for identifying and classifying both ownership and responsiveness from GitHub discussions submitted to the 500 most dependent-upon npm packages.

Applying this framework at scale, however, presents a significant challenge. Manually analyzing thousands of issues is labor-intensive and impractical at the ecosystem level. To address this, we leverage Large Language Models (LLMs) to automate key parts of the analysis. With their advanced natural language understanding capabilities, LLMs offer a promising solution for automating tasks such as bug classification, acknowledgment detection, and resolution tracking~\cite{hou2023surveylarge-llms-software-engineering}. Through this work, we aim to provide a deeper understanding of maintainer responsiveness, support downstream developers in their decision-making, and ultimately contribute to a more resilient and maintainable software ecosystem.

This study aims to answer the following research questions:
\begin{itemize}

    \item \textbf{RQ1:} \textit{To what extent do npm package maintainers respond to bug reports submitted by downstream developers?} \\
    Given that a single package may be a dependency for thousands of others, even a minor bug can have far-reaching consequences. By quantifying maintainer responsiveness, we aim to provide insights that help downstream developers make informed decisions when selecting dependencies, while also identifying packages that may require additional support and intervention to enhance their quality and long-term maintainability.
    \item \textbf{RQ2:} \textit{What are the reasons upstream developers of npm packages do not fix reported bugs?}\\
    Understanding why certain bugs remain unresolved is essential for improving collaboration between upstream and downstream developers. Identifying the specific reasons why upstream developers choose not to fix reported bugs provides critical insights into decision-making processes that affect the reliability of npm packages.
    \item \textbf{RQ3:} \textit{How can Large Language Models be used to automate the analysis of maintainer responsiveness in individual npm packages?}\\
    Manual classification of maintainer responsiveness does not scale across thousands of issues and packages. This research extends RQ1 by using instruction-tuned LLMs to automate and replicate manual classifications at the individual package level, enhancing external validity and practical applicability.



\end{itemize}


Our findings indicate that npm upstream developers are generally responsive, with a median project-level
responsiveness of 70\% (IQR: 55\%–89\%), reflecting their commitment to support downstream developers. The reasons why maintainers do not resolve these upstream bugs were categorized into four main groups: \textit{Contribution Practices} (e.g., template violations, insufficient information, and pending for proper fix), \textit{Dependency Issues} (e.g., bugs in dependencies and version incompatibilities), \textit{Library Standards} (e.g., rare edge cases, design concerns, priority, and discontinue maintenance), \textit{Lack of Engagement} (e.g., lack of comments and staled investigations). By identifying these root causes of non-responsiveness, our study offers practical insights to improve bug resolution workflows such as enhancing contribution practices, leveraging LLMs for issue triage, and fostering more sustainable collaboration between upstream and downstream developers.


The data and scripts to reproduce our results are available in the replication package.\footnote{\url{https://osf.io/v4acz}}


\section{Related Works} \label{sec:relatedWorks}
\subsection{Quality and Responsiveness in Open-Source Projects}

A well-maintained OSS project is often characterized by active community engagement, timely issue resolution, and robust code quality. Garousi~\cite{garousi2009investigating} proposed metrics such as average fix time and issue resolution rates to quantify project health. Similarly, Zhao et al.~\cite{zhao2021evaluation} and Dey and Mockus~\cite{Dey:ESEM2020} highlighted the importance of developer activity and community dynamics in sustaining quality over time.

From a security perspective, Chinthanet et al.~\cite{chinthanet2021lags} and Decan et al.~\cite{decan2018impact} emphasized that vulnerabilities in upstream packages can severely affect dependent systems. Collectively, these studies advocate for proactive maintenance, rapid patching, and strong community practices as essential factors for ensuring long-term OSS reliability.

\subsection{Upstream–Downstream Collaboration and Bug Resolution}
Several studies examine the interactions between downstream users and upstream developers in the context of bug resolution. Lin et al.\cite{lin2022upstream} found that only 13.9\% of high-severity bugs in Debian and Fedora were fixed by upstream developers, while 13.3\% were resolved by distribution maintainers, with upstream fixes generally requiring longer waiting times and more supplementary information. Their results highlight that clear references and high textual similarity between downstream and upstream reports significantly increase the likelihood of an upstream fix. Complementary findings by Zhang et al.\cite{lin2022upstream} in the Rails ecosystem show that cross-project linking, while associated with longer discussion threads, does not necessarily delay issue resolution. Additionally, Bogart et al.~\cite{bogart2021and} observed that maintenance practices differ across ecosystems such as Eclipse, npm, and CRAN, where policies around breaking changes affect how maintenance are distributed between upstream and downstream developers.

\subsection{LLMs for Issue Classification and Analysis}
Recent studies have explored the use of pre-trained language models to automate issue tracking tasks. Heo et al.\cite{heo2024comparison} introduced IssueBERT, a BERT model trained on GitHub issues, which outperformed RoBERTa by 3.6\% in F1 score on issue-label prediction. Similarly, Ardimento and Mele\cite{ardimento2020using} applied BERT for predicting bug-fix times based on bug descriptions and comments. In industry, Microsoft’s COMET system, based on GPT-3.5, improved incident triage accuracy by 30\% and reduced mitigation time by 35\%\cite{wang2024large}.

While these results demonstrate the promise of LLMs for software issue analysis, they also highlight limitations. Effective performance often requires domain-specific fine-tuning~\cite{heo2024comparison}, and input-size constraints necessitate data selection strategies~\cite{wang2024large}. Moreover, models like COMET operate as black boxes, raising concerns about interpretability. Overall, transformer models can significantly improve issue classification and triage, but they still require careful adaptation and human oversight.

Although these studies provide valuable insights, they primarily consider all bug reports without distinguishing the ownership of the bugs, and rely solely on issue status without accounting for the actual resolution or activity associated with the issues.

\section{Empirical Design} \label{sec:methodology}

\begin{figure*}
    \includegraphics[width=\textwidth]{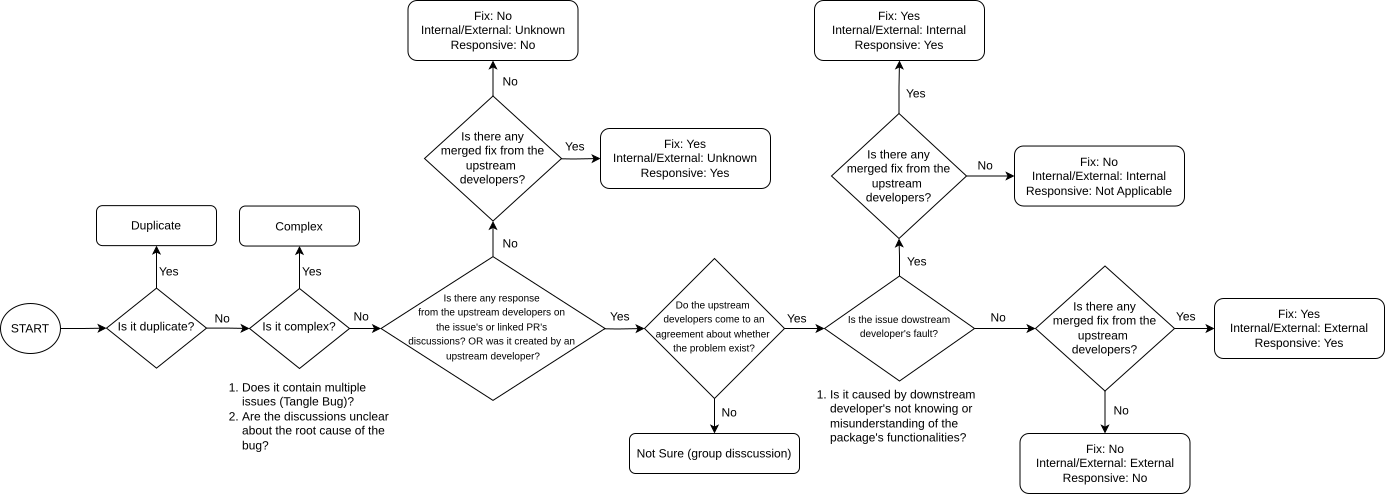}
    \centering
    \caption{Our designed framework for manual analysis of issues for responsiveness}
    \label{fig:coding-book}
\end{figure*}

\subsection{RQ1: To what extent do npm package maintainers respond to bug reports submitted by downstream developers?} \label{subsec:rq1}

To address RQ1, we develop a classification framework to label bug reports as either Responsive or Not-responsive, taking their origin into account. We build on the dataset published by Wattanakriengkrai et al.~\cite{Wattanakriengkrai2023}, applying additional filtering and performing a manual qualitative analysis to ensure the quality of the results.

\subsubsection{Data Preparation} 
\label{subsubsec:dataCollection}
Wattanakriengkrai et al. \cite{Wattanakriengkrai2023}'s dataset contains 67,212 records, including 39,320 issues and 27,892 PRs submitted to the 500 most dependent-upon npm packages up to August 2020. To prepare the data and focus only on bug reports, we apply the following filtering criteria:

Excluding Pull Requests: Bugs are typically reported by creating an issue in the corresponding GitHub repository, which initiates a discussion about their root causes. Developers often fixed these issues in a separate branch, which is then merged into the main branch via a pull request \cite{github2023pr}. Since PRs are intended to provide fixes rather than report bugs, we exclude them from our dataset, resulting in a total of 39,320 issues.
    
Identifying Bug Reports: This study aims to analyze only bug reports. Therefore, only issues explicitly reporting a bug are retained for analysis. Bug-related issues are identified based on labels assigned to issues in GitHub, typically by upstream developers. Issues labeled with the term ``Bug" and its variants (e.g., ``Bugs", ``Possibly Bug") are identified using Regular Expressions (RegEx). After applying this filter, 4,061 issues are retained.

Excluding Open Issues: Open issues are excluded under the assumption that closed issues with a ``Bug" label underwent scrutiny and confirmation by upstream developers. This filtering step reduces the dataset to 3,475 issues.

Limiting Time Frame: To focus on recent bugs, the time frame is limited between August 2017 to August 2020, yielding a final set of 1,729 issues from 83 popular npm packages for analysis.

\subsubsection{Proposed Framework}
Based on the ownership of the bug, we classify it as Responsive, Not-responsive and Not-applicable to responsiveness. It is important to note that responsiveness in our research differs from bug ownership. Bug ownership refers to who is ultimately responsible for the issue, whether it is the upstream developers or the downstream developers. For example, downstream developers may misuse a package, misunderstand its intended functionality, or fail to specify requirements correctly. In such cases, the bug is considered to be owned by the downstream developers. However, upstream developers can still choose to be responsive to these reports, even when the issue does not stem from a flaw in their code. To account for such scenarios, we also determine whether the issue arises from a misunderstanding by the downstream developers or from an actual defect in the upstream package. Based on this distinction, and a detailed examination of issue titles, descriptions and all associated discussions, we classify bugs from the perspective of downstream developers into:

\begin{itemize}
    \item \textbf{External bugs:} These occur when the downstream developers use the package correctly, and the issue lies within the package's source code or one of its dependencies. In this case, the bug is owned by the upstream project.

    \item \textbf{Internal bugs:} These arise due to incorrect usage, misunderstanding, or misinterpretation of the package by the downstream developers. Thus, the bug is considered to be owned by the downstream team.

   \item \textbf{Unknown bugs:} If there are no comments from upstream developers in the issue discussion or related PRs, it becomes impossible to determine bug ownership. Even when upstream developers fix the issue, we cannot assume they own the bug, as such fixes may be aimed at improving compatibility with downstream developers rather than addressing a defect in their own code.
\end{itemize}

If an issue was addressed—based on evidence such as a merged pull request, a submitted commit, or a comment from upstream developers—it is labeled as Responsive, regardless of whether it was classified as External, Internal, or Unknown, reflecting that upstream developers were responsive to issue. This includes cases where Internal bugs, although not caused by the upstream package, were fixed to improve compatibility with downstream users.

If the issue was not fixed, we consider bug ownership to determine the appropriate label. Internal bugs are classified as Not-applicable to responsiveness, as they often result from downstream errors and do not necessarily require upstream intervention. In contrast, unresolved External or Unknown bugs are classified as Not-responsive.

In addition, the framework also identifies duplicate and complex issues, which were excluded from the analysis, as their inclusion would introduce more noise than meaningful insights:

\begin{enumerate}
    \item {Duplicate Issues: }
    There may be multiple issues submitted to GitHub repositories describing the same bug. Upstream developers often mark such duplicate issues within the issue's discussions. Since updates related to acknowledgment or resolution appear only on the original issue, duplicate issues lack relevant follow-up information and are excluded from further analysis.
    
    \item {Complex Issues:} 
    An issue is labeled as complex if it contains multiple unrelated problems (tangled bugs), unrelated discussions, or if the discussions are too ambiguous for a reviewer to make a clear decision. If an issue is classified as complex, we do not proceed with its classification.\footnote{An example of a complex issue: \url{https://github.com/jsx-eslint/eslint-plugin-react/issues/2570}}
\end{enumerate}
We created a decision diagram based on the classification framework (Figure~\ref{fig:coding-book}) to guide participants during the manual analysis.

\subsubsection{Manual classification}
Seven participants—two professors and five master’s students in Computer Science—participate in the classification process. All participants have prior experience with JavaScript. To ensure clarity and consistency, we conduct multiple rounds of discussion to align all participants with the proposed classification framework. Each researcher independently analyzes the same subset of issues, and their results are reviewed in follow-up meetings. Disagreements are discussed and resolved collaboratively. This iterative process continues until a shared understanding of the framework and its application to edge cases is established, ensuring consistency before proceeding with the full individual classification.

For the full classification task, we create a spreadsheet containing URLs of the 1,729 issues and distribute it among the participants. The spreadsheet includes the following columns: IsDuplicate (Yes or No), IsComplex (Yes or No), Type (Internal or External), and WasFixed (Yes or No). Participants review the titles, descriptions, and all associated discussions for each issue and fill in these columns. If a participant is uncertain about how to classify a specific issue, they mark it for group discussion to ensure consistency and accuracy. The Responsiveness column is subsequently inferred based on the values of the other four columns.

To evaluate the accuracy of the manual classification process, two of the authors review a random sample of 120 issues after classification is completed. During this review, only seven issues (5.8\%) are found to be misclassified and are corrected.

\subsection{RQ2: What are the reasons upstream developers of npm packages do not fix reported bugs?} \label{subsec:rq2}
After classifying issues based on responsiveness, we retain all Not-responsive bug reports (i.e., unfixed External and Unknown bugs), resulting in a total of 87 issues. We then conduct an open coding process to develop a taxonomy of the reasons behind unresolved bugs in the npm ecosystem.

Two researchers participate in the open coding process. The 87 issues are divided into groups of 20 per round (with the final round containing 7 issues). In each round, both researchers independently review the issue titles and discussions to identify and summarize the reasons why the bugs remain unresolved. During this process, they generate preliminary labels. After each round, the researchers meet to discuss and align on the labels, merging similar ones and reclassifying issues when new labels emerge. Finally, the labels are consolidated into broader categories, resulting in a taxonomy of underlying reasons for unresolved bug reports in the 500 most depended-upon npm packages.

\subsection{RQ3:  How can Large Language Models be used to automate the analysis of maintainer responsiveness in individual npm packages?} \label{subsec:rq3}
In our analysis for RQ1, we were unable to manually evaluate all issues submitted to each of the 500 packages individually. Instead, we analyzed maintainer responsiveness across the combined set of packages. In RQ2, our goal is to evaluate responsiveness at the level of each individual package. To achieve this, we develop a pipeline where instruction-tuned LLMs first distinguish bug reports from other types of issues and then classify each bug report as either Responsive or Not-responsive across the top 500 npm packages. We selected three LLMs for our experiments, each with 8 billion parameters: Llama 3 Instruct and Llama 3.1 Instruct, both released in 2024 \cite{grattafiori2024llama}, and DeepSeek R1, a distilled variant of Llama introduced in 2025~\cite{guo2025deepseek}. These models are chosen for their strong performance relative to size, and accessibility through smaller variants that can run on limited hardware.

\subsubsection{Bug Report Identification} \label{subsubsec:bug-report-iden}
As described in Section \ref{subsec:rq1}, while labels like ``Bug" help identify bug reports, inconsistencies across projects may limit their reliability. To address this, we used a two-step approach: (1) create a ground truth dataset of bug and non-bug issues, and (2) employ LLMs to classify issues using contextual information. This enables accurate large-scale bug report identification across 500 packages, supporting deeper analysis of reporting and resolution trends.

\textbf{Ground Truth Dataset Curation:}
To evaluate LLM performance, we use the curated dataset from Section~\ref{subsec:rq1}, which contains 1,729 bug report issues from 83 packages. To create a balanced benchmark, we augment this dataset with non-bug issues. To identify such issues, we first extract all labels used across the 83 packages. We then exclude known bug labels (identified in RQ1) as well as any labels that had ever co-occurred with a bug label, ensuring the remaining labels are unrelated to bugs. This process yields 734 unique non-bug label variations.

To further refine the selection, one of the authors initially removes labels that may still be related to bugs (e.g., ``Help Wanted," ``Security," ``Need Investigation"). Then, two authors review the remaining labels and select 10 (comprising 94 variations). For each label, they sample five issues and eliminate any that still appear to be bug reports. Ultimately, five labels (with 45 variations) are chosen to represent non-bug issues: ``Feature," ``Request," ``Idea," ``Proposal," and ``Quality." Issues labeled exclusively with these are treated as non-bug issues. This process results in 217 candidate non-bug issues. After a final manual review, 19 misclassified bug reports are removed.

\begin{figure}
    \includegraphics[width=\columnwidth]{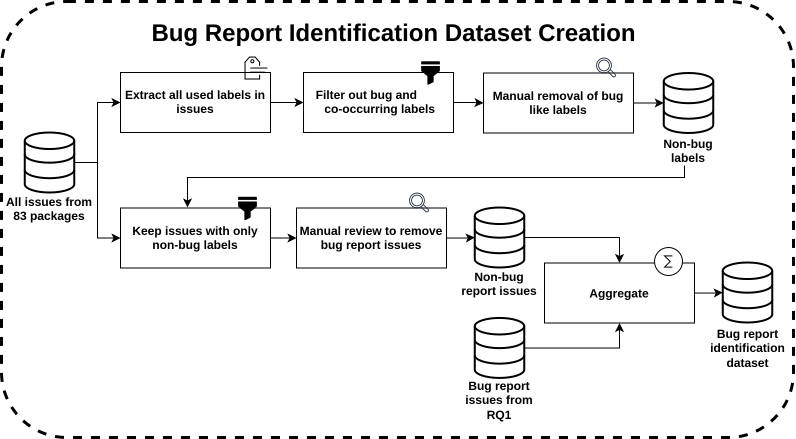}
    \centering
    \caption{Overview of the sequence of steps followed to construct the dataset for bug report identification.}
    \label{fig:bug-iden-dataset-creation}
\end{figure}

The resulting dataset contains 1,927 issues, including 1,729 bug reports and 198 non-bug issues. Figure~\ref{fig:bug-iden-dataset-creation} illustrates the sequence of steps used to construct this dataset.

\textbf{Context Generation:}
In this section, we employed instruction-tuned LLMs in a zero-shot setting. Zero-shot learning with LLMs refers to leveraging a model's pretrained knowledge and natural-language instructions—without any task-specific fine-tuning—to perform a new task~\cite{kojima2022large, mann2020language, roumeliotis2025llms, qiu2024chatgpt}. In our case, the model receives only an instruction and the context generated for each issue, which includes the issue title and its description. This approach enables highly scalable classification, as it requires little to no labeled training data.

\textbf{Evaluating the LLMs for Bug classification:} 
The evaluation was conducted on the curated ground truth dataset. Each model was assessed using standard classification metrics: accuracy, precision, recall, and F1 score. To enhance results, we iteratively tested different instructions and evaluated their impact on classification performance. The final evaluation used the best-performing instruction (available in our replication package). In this instruction, the model is instructed to generate only a \textit{Json} object containing a single field, whose definition is also provided.

To further optimize performance, we explored different hyperparameter configurations by varying the temperature (0.0, 0.2, 0.5, 0.8, 1.0) and Top-p (0.7, 0.8, 0.9, 0.95) values. Each combination was evaluated using the same set of classification metrics to identify the most effective setup.

\textbf{Classification Phase:} 
Wattanakriengkrai et al. \cite{wattanakriengkrai2020github}'s dataset includes issues up to August 2020. We extended it using the GitHub API \cite{githubAPIIssue} to include all issues from the same 500 packages up to January 2025, resulting in 103,082 entries. After filtering as described in Section \ref{subsec:rq1}, we removed 47,741 PRs and excluded 7,458 open issues, yielding a dataset of 47,883 closed issues. Using the best-performing model and the most effective hyperparameter configuration, we classified all 47,883 issues as either bug reports or non-bug issues.

\subsubsection{Responsiveness Classification}
We used the issues classified as bug reports from the previous step and enriched the dataset by collecting event data from issue timelines using the GitHub API \cite{githubAPITimeLine}. We also retrieved the list of contributors using the GitHub API \cite{githubAPIContributors} to distinguish between upstream and downstream interactions, which is critical for accurately modeling issue resolution and acknowledgment processes.

Then, we classified bug reports as Responsive and Not-responsive using an instruction-tuned LLM in a zero-shot setting. This approach follows a similar classification process as that performed manually in Section \ref{subsec:rq2}. To ensure the accuracy and reliability of the LLM’s classification, we used the results from our manual analysis (Section \ref{subsec:rq2}) as ground truth to evaluate the model’s performance. We systematically refined the input context and instructions to maximize the model’s classification accuracy. 

\textbf{Context Generation:} 
In this task, the context for each issue includes issue metadata such as title, description, and selected activity events, which we parse and structure into a single text block to mimic the context available during manual analysis. 

While the GitHub API~\cite{githubAPIEventTypes} defines 47 possible issue event types, not all are relevant for our classification task. Including irrelevant events may introduce noise and unnecessarily increase input length, potentially degrading model performance~\cite{wang2024large}. Drawing on insights from our manual analysis, we select 13 key event types that offer meaningful signals related to bug acknowledgment, resolution, and duplicate detection.

By carefully selecting and filtering relevant events, we constructed high-quality input representations that capture the issue lifecycle and interactions between upstream and downstream developers. This refinement improved the LLM’s ability to replicate human-level classification accuracy and ensured that the resulting dataset is both comprehensive and reliable.

\textbf{Evaluating LLMs:}
Before using LLMs for responsiveness classification, we conducted a series of experiments on the manually labeled ground truth dataset from Section~\ref{subsec:rq1}. Our goal was to identify the best-performing model, optimize hyperparameters, and refine the instruction to maximize classification accuracy. During the prompt engineering phase, we tested 13 different instructions. The best one we selected is available in our replication package.

\textbf{Classification Phase:} Finally, we classified all bug reports identified in stage one into duplicate, Responsive, Not-responsive, or Not-applicable to responsiveness categories, using the best-performing model, hyperparameters, and instruction.

\section{Results} \label{sec:results}

\subsection{RQ1: Responsiveness of npm package maintainers \label{subsec:results-rq1}}
Using the proposed framework, we analyzed 1,729 issues. Among these, 64 issues (4\%) were identified as duplicates and 47 (3\%) as complex. The remaining 1,618 issues (93\%) were classified as either Internal, External, or Unknown bugs. Table \ref{tab:res_manual_classification} summarizes these results. Specifically, 89 issues (6\%) were categorized as Internal bugs, 1,426 (88\%) as External bugs, and 103 (6\%) as Unknown. In total, we identified 1,458 (90\%) issues as Responsive, 87 (5\%) as Not-responsive, and 73 (5\%) as Not-applicable in terms of responsiveness. In terms of responsiveness, we report three findings.

First, among the External bugs, we found 42 (3\%) Not-responsive issues. This indicates that although upstream developers acknowledged that the bugs originated from the packages' source code, they either did not want to or could not fix them. However, in 1,384 (97\%) issues of External bugs, they were Responsive, meaning that they addressed the needs of their downstream developers by providing fixes.

Second, among the 89 issues classified as Internal bugs, we found 16 (18\%) Responsive issues. This indicates that although the bugs were in downstream developers' source code, upstream developers attempted to provide fixes to better align with their downstream developers' needs. Additionally, 73 (82\%) issues were classified as Not-Applicable, meaning that they were not relevant to responsiveness. These bugs originated from downstream developers' source code, and upstream developers were unable to fix them.

Third, among the 103 issues classified as Unknown, meaning that there were no discussions from upstream developers, we found 58 (56\%) Responsive issues. This suggests that although upstream developers did not participate in discussions submitted to the issues or PRs, they still provided fixes for the reported bugs. On the other hand, we found 45 (44\%) Not-responsive issues, indicating that upstream developers neither participated in the discussions nor provided any fix for the reported bugs.

\begin{table}
    \centering
    \caption{Distribution of Internal, External, and Unknown bugs as well as Responsive, Not-responsive, and Not-applicable issues in the ground truth dataset.}
    \label{tab:res_manual_classification}
    \begin{tabular}{lcccc}
         \toprule
         & \textbf{Internal} & \textbf{External} & \textbf{Unknown} & \textbf{Total} \\
         \toprule
         \textbf{Responsive} & 16 (18\%) & 1,384 (97\%) & 58 (56\%) & 1,458 (90\%) \\
         \textbf{Not-responsive} & - & 42 (3\%) & 45 (44\%) & 87 (5\%) \\
         \textbf{Not-applicable} & 73 (82\%) & - & - & 73 (5\%) \\
         \midrule
         \textbf{Total} & 89 (6\%) & 1,426 (88\%) & 103 (6\%) & 1,618 \\
         \bottomrule
    \end{tabular}
\end{table}

\begin{tcolorbox}
    \textit{\textbf{RQ1:} npm package maintainer responsiveness}

    \textit{\textbf{Summary of results:}} 
    Overall, 90\% of bug reports were classified as Responsive, suggesting that upstream developers are generally attentive to downstream developers' needs. We also found that the majority of bug reports (88\%) are External, indicating that they stem from actual defects in the package.
\end{tcolorbox}

\subsection{RQ2: Taxonomy for Non-responsiveness \label{subsec:results-rq2}}
We categorized the reasons for not resolving bugs into four main categories, each with subcategories. The occurrences of these reasons are also summarized in Table \ref{tab:taxonomy}.

\begin{table}
    \centering
    \caption{Main reasons for not addressing bugs in the npm ecosystem, along with their occurrence.} \label{tab:taxonomy}
    \begin{tabular}{llc}
        \toprule
        \textbf{Category} & \textbf{Sub-Category} & \textbf{\# Occurrence} \\
        \toprule

        \multirow{3}{*}{Contribution Practices} & Template violation & 34 (39\%) \\
        & Insufficient information & 6 (7\%) \\
        & Waiting for pull request & 5 (6\%) \\
        \midrule
        
        \multirow{3}{*}{Dependency} & Dependency issue & 13 (15\%) \\
        & Incompatible versions & 3 (3\%) \\
        & Beyond the scope  & 2 (2\%) \\
        \midrule

        \multirow{4}{*}{Library Standards} & Edge case & 3 (3\%) \\
        & Design concern & 2 (2\%) \\
        & Priority & 1 (1\%) \\
        & Discontinue maintenance & 1 (1\%) \\
        \midrule

        \multirow{2}{*}{Lack of Engagement} & No comments & 5 (6\%) \\
        & Staled investigation & 12 (14\%) \\
        
        \bottomrule
    \end{tabular}
\end{table}

\textbf{1. Contribution Practices}: refers to cases where bugs remain unresolved due to how contributions are managed.
\begin{itemize}
    \item \textit{Template violation:} Although the package provides a predefined bug report template, the reporter did not use it. As a result, the issue was not fixed and closed.

    \item \textit{Insufficient Information:} Upstream developers often attempt to reproduce reported bugs by replicating the conditions described. However, when they are aware of a bug but cannot reproduce it, they typically request additional details from the reporter. If the reporter fails to provide the needed information, the issue is closed and not fixed until it is reported again with sufficient context.

    \item \textit{Waiting for pull request:} The root cause of the bug has been identified, but a proper solution has yet to be found. The team is awaiting contributions from other upstream developers to propose a viable fix.
\end{itemize}

\textbf{2. Dependency}: Not all bugs originated from a project's own code; some can arise from third-party dependencies \cite{rodriguez2020bugs,rodriguez2018if}. These issues often remain unresolved as they are outside the immediate control of the maintainers.

\begin{itemize}
    \item \textit{Dependency issue:} The bug originates from one of the package's dependencies, and the fix should be applied by that dependency. Once the dependency library resolves the issue, the bug can be resolved by upgrading to the latest version of the dependency.

    \item \textit{Incompatible versions:} The bug is caused by one of the package's dependencies, which was fixed in a newer version of the dependency. However, the package is incompatible with this updated version, possibly due to breaking changes, preventing an upgrade to the fixed version.

    \item \textit{Beyond the scope:} The bug is caused by one of the package's dependencies, but the dependency has not acknowledged the bug and will not provide a fix. As a result, resolving this bug is beyond the control of the package.
\end{itemize}

\textbf{3. Library Standards:} refers to issues arising from a package's adherence to best practices and conventions.
\begin{itemize}
    \item \textit{Edge case:} This bug occurs only in very rare circumstances, and the package has decided not to allocate resources toward fixing it.

    \item \textit{Design concern:} Although the bug exists in the package, it only arises when downstream developers fail to follow best practices.

    \item \textit{Priority:} Fixing this bug is not feasible at the moment, as it would require major refactoring or risk negatively impacting other parts of the package.

    \item \textit{Discontinue maintenance:} The package is no longer maintained.
\end{itemize}

\textbf{4. Lack of Engagement:} refers to issues where upstream developers fail to actively address or respond to reported bugs.
\begin{itemize}
    \item \textit{No comments:} The issue was closed without any response/comments from the upstream developers. 
    
    \item \textit{Staled investigation:} The upstream developers acknowledged the bug but eventually stopped investigating it and closed the issue without fixing it.
\end{itemize}

\begin{tcolorbox}
    \textit{\textbf{RQ2:} Reasons for not being responsive}

    \textit{\textbf{Summary of results:}} 
    We identified four main reasons why bugs were not resolved in the npm ecosystem. The most common was Contribution Practices, accounting for 52\% of all Not-responsive bug reports, followed by Dependency issues (20\%), Lack of Engagement (20\%), and Library Standards (7\%).
      
\end{tcolorbox}

\subsection{RQ3: Automated Classification} 
We divided our results into: bug report identification and responsiveness classification. Table \ref{tab:bug_iden_results} presents the results and highlights the best-performing model for bug report identification. The performance of each configuration was evaluated using accuracy, precision, recall, and F1-score. 

Our findings indicate that all models achieved a precision of 1.0, meaning every issue classified as a bug report was indeed a true bug report, thereby eliminating false positives. However, other performance metrics varied.

DeepSeek R1 achieved the highest overall accuracy of 0.94 and the highest recall of 0.93, resulting in an F1-score of 0.96, which matched Llama 3. This indicates that DeepSeek R1 was the most effective at correctly identifying the majority of actual bug reports without compromising on precision. Llama 3 also achieved an F1-score of 0.96, with a slightly lower recall (0.92) and overall accuracy (0.93). In contrast, Llama 3.1 recorded the lowest accuracy (0.91) and recall (0.90), leading to a slightly lower F1-score (0.95).

While all models demonstrate high reliability in terms of precision, DeepSeek R1 offered the most balanced performance by effectively maximizing both recall and overall accuracy. This balance is especially important for the next classification task, where it is crucial not only to avoid false positives but also to minimize false negatives, that is, missed bug reports. Therefore, we selected DeepSeek R1 as the preferred model for classifying all issues submitted to the 500 selected packages.

\begin{table*}[ht]
    \centering 
    \caption{Performance evaluation of LLMs for bug report identification, including the best-performing Temperature and Top-p values.} \label{tab:bug_iden_results} 
        \begin{tabular}{lcccccc} 
            \toprule 
            \textbf{Model Name} & \textbf{Temperature} & \textbf{Top-p} & \textbf{Accuracy} & \textbf{Precision} & \textbf{Recall} & \textbf{F1-score}\\
            \toprule 
            Llama 3 Instruct & 0.5 & 0.9 & 0.93 & \textbf{1.00} & 0.92 & \textbf{0.96} \\ 
            Llama 3.1 Instruct & 0.2 & 0.95 & 0.91 & \textbf{1.00} & 0.90 & 0.95 \\ 
            \textbf{DeepSeek R1 (Distill Llama)} & 0.2 & 0.9 & \textbf{0.94} & \textbf{1.00} & \textbf{0.93} & \textbf{0.96} \\ 
            \bottomrule 
        \end{tabular} 
\end{table*}

\begin{table*}[ht]
    \centering
    \caption{Performance evaluation of LLMs for responsiveness classification, including the best-performing Temperature and Top-p values.}
    \label{tab:res_classification_results}
        \begin{tabular}{l|ccc|ccc|ccc}
            \toprule
            \multirow{4}{*}{\textbf{Class}} & \multicolumn{3}{c|}{\textbf{Llama 3 Instruct}} & \multicolumn{3}{c|}{\textbf{Llama 3.1 Instruct}} & \multicolumn{3}{c}{\textbf{DeepSeek R1 (Distill Llama)}} \\
                           & \multicolumn{3}{c|}{Temperature: 0.2} & \multicolumn{3}{c|}{Temperature: 0.5} & \multicolumn{3}{c}{Temperature: 0.2} \\
                           & \multicolumn{3}{c|}{Top-p: 0.7} & \multicolumn{3}{c|}{Top-p: 0.8} & \multicolumn{3}{c}{Top-p: 0.7} \\
            \cline{2-10}
                           & \textbf{Precision} & \textbf{Recall} & \textbf{F1-score} & \textbf{Precision} & \textbf{Recall} & \textbf{F1-score} & \textbf{Precision} & \textbf{Recall} & \textbf{F1-score} \\
            \hline
            Duplicate      & 0.23 & 0.55 & 0.32 & 0.25 & \textbf{0.56} & 0.35 & \textbf{0.31} & 0.55 & \textbf{0.39} \\
            Responsive     & 0.94 & \textbf{0.90} & \textbf{0.92} & 0.95 & 0.89 & \textbf{0.92} & \textbf{0.96} & 0.75 & 0.84 \\
            Not-responsive & \textbf{0.46} & \textbf{0.67} & \textbf{0.54} & 0.33 & 0.51 & 0.40 & 0.12 & 0.47 & 0.19 \\
            Not-applicable & \textbf{1.00} & 0.03 & 0.05 & 0.19 & 0.08 & 0.12 & 0.18 & \textbf{0.23} & \textbf{0.21} \\
            \hline
            \textbf{Accuracy} & \multicolumn{3}{c|}{\textbf{0.84}} & \multicolumn{3}{c|}{0.82} & \multicolumn{3}{c}{0.71} \\
            \bottomrule
        \end{tabular}
\end{table*}

Table~\ref{tab:res_classification_results} shows the results of the responsiveness classification evaluated using standard metrics: precision, recall, F1 score, and accuracy. 

All models performed well in Responsive class, with Llama 3 Instruct and Llama 3.1 Instruct both reaching F1-scores of 0.92, indicating their strong ability to detect issues that were addressed. However, DeepSeek R1 underperformed in this category, with a lower recall (0.75) and a resulting F1-score of 0.84, which may lead to underestimating the number of issues that actually received responses. Llama 3's high precision (0.94) and recall (0.90) in this class affirm its reliability and consistency. In the Not-responsive category, Llama 3 Instruct demonstrated the strongest performance. It achieved the highest F1-score of 0.54, significantly outperforming Llama 3.1 (F1 = 0.40) and DeepSeek R1 (F1 = 0.19). This result highlights Llama 3's superior balance between precision (0.46) and recall (0.66) in identifying unresolved bug reports.  

For the auxiliary Duplicate class, DeepSeek R1 achieved the highest F1-score (0.39), with a strong precision of 0.31 and recall of 0.55, outperforming both Llama variants. Although Llama 3.1 had the highest recall (0.56), its lower precision (0.25) reduced its F1-score (0.35). Llama 3 Instruct trailed slightly with an F1 of 0.32. In the Not-applicable class, DeepSeek R1 again provided the best performance with an F1-score of 0.21, owing to its relatively high recall (0.23). In contrast, Llama 3 struggled in this class, despite having the highest precision (1.00), because of an extremely low recall (0.03), resulting in a much lower F1-score of 0.05. 

When considering overall performance, Llama 3 Instruct achieved the highest accuracy of 0.84, compared to 0.82 for Llama 3.1 and 0.71 for DeepSeek R1. This metric supports the conclusion that Llama 3 made fewer incorrect predictions across all classes. While accuracy alone is not sufficient in imbalanced or multi-class settings, in combination with strong F1-scores on the two most important classes, it strengthens the case for Llama 3 Instruct as the most reliable and consistent model overall.

Although DeepSeek R1 occasionally outperformed the Llama models, particularly in the Duplicate and Not-applicable classes—these are auxiliary categories in our analysis. Its lower F1 in the Not-responsive and Responsive classes, makes it less suitable for this RQ. Llama 3.1 Instruct did well on some metrics but scored lower than Llama 3 on Not-responsive and only matched it on Responsive.
Llama 3 Instruct was selected as the final model due to its superior performance on the Not-responsive class and its equally strong results on the Responsive class. No other model approached its F1-score of 0.54 for Not-responsive issues. 

Using Llama 3, we analyzed the 30,340 issues previously classified as bug reports in Section~\ref{subsubsec:bug-report-iden}. For 24 of these issues, the model failed to generate a meaningful or structured output, and they were excluded from further analysis. Among the remaining issues, 3,244 (11\%) were marked as duplicates and similarly excluded. The final set of 27,072 (89\%) bug reports was classified based on bug type into: \textbf{Internal}, \textbf{External}, and \textbf{Unknown}.

\begin{table}[b]
    \centering
    \caption{Distribution of Internal, External, Unknown, Responsive, Not-responsive, and Not-applicable issues. Each cell reports the absolute number of issues, with the percentage in parentheses indicating the median ratio across projects.} 
    \begin{adjustbox}{width=0.5\textwidth}
        \begin{tabular}{lcccc}
             \toprule
             & \textbf{Internal} & \textbf{External} & \textbf{Unknown} & \textbf{Total} \\
             \toprule
             \textbf{Responsive} & 440 (33\%) & 15,830 (78\%) & 673 (20\%) & 16,943 (70\%) \\
             \textbf{Not-responsive} & - & 6,370 (22\%) & 2,944 (80\%) & 9,314 (25\%) \\
             \textbf{Not-applicable} & 815 (67\%) & - & - & 815 (0\%) \\
             \midrule
             \textbf{Total} & 1,255 (0\%) & 22,200 (91\%) & 3,617 (0\%) & 27,072 \\
             \bottomrule
        \end{tabular}
    \end{adjustbox}
    \label{tab:res_autamated_classification}
\end{table}

Table~\ref{tab:res_autamated_classification} presents the distribution of issues across three bug types (Internal, External, and Unknown) and three responsiveness categories (Responsive, Not-responsive, and Not-applicable). The absolute number of issues is shown in each cell, and the percentage in parentheses represents the median ratio specific to that category. 

External bugs account for the majority of the dataset, with 22,200 issues, followed by 3,617 unknown bugs and 1,255 internal bugs. The median share of external bugs per project is 91\% (IQR: 79\%–100\%), meaning that in half of the projects, more than 91\% of reported bugs are external. Internal bugs represent a much smaller proportion, with a median of 0\% (IQR: 0\%–7\%), indicating that many projects either had no internal bug reports or very few. Similarly, unknown bugs show a median of 0\% (IQR: 0\%–12\%), suggesting that bugs with unclear origin are relatively rare across most projects.

In terms of responsiveness, external bugs were more actively addressed. Specifically, 15,830 external bugs were classified as Responsive, with a median responsiveness rate of 78\% (IQR: 64\%–100\%), meaning that alf of the projects resolved more than 78\% of their external bugs. The median not-responsiveness rate for external bugs is 22\% (IQR: 0\%–36\%), showing that although the majority of external bugs were addressed, some projects still had a significant backlog of unresolved external bugs.

Internal bugs exhibit a distinct pattern. Only 440 internal bugs were Responsive, with a much lower median responsiveness rate of 33\% (IQR: 0\%–53\%). In addition, a large proportion of internal bugs (815 cases) were classified as Not-applicable to responsiveness evaluation, with a median not-applicability rate of 67\% (IQR: 47\%–100\%). 

Unknown bugs are fewer in number but show the highest rates of neglect. Only 673 unknown bugs were resolved, yielding a median responsiveness rate of 20\% (IQR: 0\%–50\%). Conversely, 2,944 unknown bugs remained unresolved, with a high median not-responsiveness rate of 80\% (IQR: 50\%–100\%).

Aggregating across all issue types, 16,943 issues were classified as Responsive overall, with a median project-level responsiveness of 70\% (IQR: 55\%–89\%). The median not-responsiveness rate across all packages was 25\% (IQR: 0\%–40\%), while issues classified as Not-applicable represented a small fraction of the dataset, showing a median of 0\% (IQR: 0\%–3\%). In summary, external bugs dominate in both volume and responsiveness, internal bugs are frequently Not-applicable to responsiveness evaluation, and unknown bugs are often left unresolved.

The marked gap between the manual and LLM-based results particularly the increase from 5\% to 34\% in the Not-responsive category, can be attributed to differences in sampling and coverage. Our manual coding covered 1,618 bug reports. 
In contrast, the automated classification was applied to all 27,072 candidate bug reports, thereby encompassing a broader range of projects—including many long-tail packages that were excluded from the manually analyzed subset. The inclusion of these lower-activity packages in the automated analysis results in a higher proportion of Not-responsive cases, reflecting the realities of less curated or less actively maintained segments of the ecosystem.

\begin{tcolorbox}
    \textit{\textbf{RQ3:} Identifying responsiveness automatically}

    \textit{\textbf{Summary of results:}}
    We developed a pipeline using instruction-tuned LLMs in zero-shot setting to automate two tasks: bug report identification and responsiveness classification. We find that: LLMs, even in zero-shot setting, are able to distinguish between bug reports and non-bug issues with high accuracy (91\%-94\%), and classify bug reports based on bug type and responsiveness with a relatively high overall accuracy (71\% - 84\%).
    Our final results revealed that package maintainers are generally responsive, with a median project-level responsiveness of 70\% (IQR: 55\%–89\%).
\end{tcolorbox}

\section{Discussion and Practical Implications} \label{sec:implications}  
Our findings provide practical insights for both upstream and downstream developers in the npm ecosystem. 

\subsection{Upstream Developer Perspective}  
Our analysis shows that upstream maintainers are generally responsive. About half of the packages have a responsiveness ratio above 70\%. This responsiveness is important because previous studies indicate that bug fixes motivate downstream developers to update their dependencies~\cite{salza2020third,derr2017keep}. However, the wide variation in responsiveness across packages suggests differences in maintenance resources or community involvement. Our manual classification identified \textit{Lack of Engagement} as the second most common reason for unresolved issues, accounting for 20\% of all Not-responsive bug reports. This often occurs when maintainers do not participate in issue discussions or abandon them prematurely due to overwhelming workloads.  

To address these gaps, we propose the following strategies:  
\subsubsection{Fostering Contributor Engagement}  
Increasing and maintaining contributor involvement can ease the burden on individual maintainers. Effective practices include:  
\begin{itemize}  
    \item Clearly outlining contribution guidelines in a \texttt{README} or \texttt{CONTRIBUTING.md} files.  
    \item Regularly updating documentation and marking outdated content to minimize confusion.  
    \item Labeling simple tasks as ``good first issue" to attract newcomers.  
    \item Promoting engaged contributors to collaborator roles for triage and support responsibilities.  
\end{itemize}  

\subsubsection{Scaling Triage with LLMs}  
Manual triage, such as categorizing issues and crafting initial responses, often leads to maintainer burnout~\cite{li2021you}. Our results show that LLMs can help reduce this burden through automation:  
\begin{itemize}  
    \item \textbf{Automated Issue Classification:} LLMs can identify bugs, feature requests, and questions, and automatically apply labels. For example, trIAge~\cite{trIAgeEBot} uses LLMs to categorizing issues, detecting duplicates, and identifying related issues.
    \item \textbf{Prioritization Support:} LLMs can evaluate bug descriptions and suggest priority levels, such as critical, high, or medium.  
    \item \textbf{Improving Report Quality:} Instruction-tuned LLMs can guide users to submit complete reports by prompting for missing details like reproduction steps or logs.  
\end{itemize}  

By incorporating these tools, maintainers can automate routine tasks like labeling, grouping, and responding to issues, allowing more time for critical development work.  

\subsubsection{Moving Beyond Temporary Workarounds}
Although our classification distinguishes between formal fixes and non-fixes, our manual review reveals that upstream maintainers often offer workarounds, suggestions, or engage in discussions instead of delivering actual fixes. While helpful, these responses are temporary and frequently undocumented, making them hard for others to discover.

Without formal fixes, downstream developers may have to search through GitHub issues for relevant solutions. This process is time-consuming and prone to error. It can lead to duplicate issue submissions, which accounted for 11\% of bug reports in our dataset, further increasing the maintainer workload. Thus, we recommend that maintainers prioritize delivering formal, documented solutions whenever possible.

\subsubsection{Mitigating Dependency-Driven Failures}
Bugs related to dependencies and version incompatibilities remain a persistent challenge. Prior research has highlighted the difficulty of managing dependencies across software ecosystems~\cite{hejderup2015dependencies, kikas2017structure, wittern2016look}. These challenges are not limited to JavaScript but are also prevalent in ecosystems like Rust and Ruby~\cite{kikas2017structure}.

Extrinsic bugs—those originating in dependencies~\cite{rodriguez2020bugs,rodriguez2018if}—can propagate through the dependency chain, impacting multiple packages. Addressing them requires coordination between upstream developers from multiple packages, better dependency management practices, and ecosystem-level tooling to trace and mitigate upstream breakages.

\subsection{Downstream Developer Perspective}
Our analysis indicates that the most common reason for non-responsiveness is low-quality bug reports from downstream developers. Specifically, 46\% of Not-responsive reports did not follow the upstream project's issue template or lacked enough information to reproduce the bug. Therefore, we strongly encourage downstream developers to carefully follow the bug reporting templates and guidelines set by upstream maintainers, and provide detailed reproduction steps, environment information, error logs, and the expected versus actual behavior. By submitting well-structured and complete bug reports, downstream developers can help close the communication gap and enable upstream teams to resolve issues more effectively.

\section{Threats to Validity} \label{sec:limitations}


\textit{Construct Validity}: 
One issue is the operationalization of \textit{responsiveness}, which is defined based on issue activity on GitHub, such as comments, labeling behavior, and closure actions. While this provides a useful measure of engagement, it may not fully capture all forms of maintainer responsiveness, such as interactions occurring off-platform or through private channels, or fixes implemented through code commits that are not explicitly tied to issues. Another challenge arises from the classification of bug ownership, which can be subjective. Determining whether a bug is the responsibility of upstream or downstream developers often depends on the specific context of the issue. While we employed a rigorous rubric to classify issues, there are cases where ownership is ambiguous, particularly when bugs result from unclear documentation or integration misuse. Additionally, the heuristics used to distinguish between bug reports and non-bug issues may not fully capture complex or ambiguous cases. Although this approach works well in straightforward scenarios, it might oversimplify the classification in more nuanced instances.

\textit{Internal Validity} 
Highly depended-upon packages tend to have more contributors and more robust maintenance practices, which may correlate with higher responsiveness. As our study focuses on the most depended-upon npm packages, we recognize that these packages may exhibit responsiveness patterns different from less popular or newly created packages. The identification of issue types presents another challenge. There may be cases where non-bug issues, such as feature requests or questions, are misclassified as bugs. 

\textit{External Validity:} 
Our study focuses exclusively on the top 500 most depended-upon npm packages. As such, our findings may not generalize to less popular, smaller, or newly published packages, which may exhibit different behaviors in terms of maintainer responsiveness. The npm ecosystem itself is also distinct, 
therefore, the findings of this study may not directly apply to other ecosystems, such as PyPI (Python), Maven (Java), or crates.io (Rust), which may have different governance structures, community norms, and tooling.

\textit{Conclusion Validity:} 
A key threat in our study is the reliance on LLMs for issue classification. Although we performed validation on a labeled subset of data, LLMs can still make misclassifications or provide unpredictable results, which could affect the robustness of our conclusions. Furthermore, the manual coding process used to classify reasons for non-resolution may introduce subjective bias, as coders could interpret issues differently based on their own perspectives or experiences. While we employed reliability checks to mitigate this risk, some subjectivity in the coding process might remain. Finally, we note that the dataset used to answer RQ1 was collected in 2020. Although the nature of maintainer responsiveness is relatively stable over short periods, more recent issue activity may differ, and this temporal gap may limit our findings.

\section{Conclusion and Future Work} \label{sec:conclusion}
In this study, we conducted a large-scale empirical analysis, investigating 30,340 issues across 500 of the most depended-upon npm packages to assess maintainers' responsiveness to bug reports, understand the reasons behind unresolved issues, and explore the feasibility of automating responsiveness classification using LLMs.

Our findings indicate that npm upstream developers are generally responsive, with a median project-level responsiveness of 70\% (IQR: 55\%–89\%), reflecting their commitment to support downstream developers. However, a notable minority of bugs remained unresolved, driven by factors such as contribution practices, dependency-related challenges, adherence to library-specific standards, and lack of engagement. By developing a taxonomy of these reasons, we provide insights that can help improve collaboration between upstream maintainers and downstream users.

Furthermore, our results demonstrate that LLMs offer a promising approach for automating responsiveness classification. Our experiments showed that LLMs can reliably distinguish between bug and non-bug issues, achieving an F1-score of 0.96, and classify responsiveness outcomes with a reasonably high accuracy of 0.84. This level of automation can substantially reduce the manual effort required for large-scale analyses of issue responsiveness in open-source ecosystems.

Future research should explore the intricate relationship between responsiveness and package reputation, particularly how various forms of engagement, such as workarounds and discussions, impact users' perceptions of popularity and quality. This includes offering a more comprehensive view of how upstream developers address bug reports. Additionally, investigating the role of automation in bug tracking and its potential biases will help in understanding how automated actions affect the perceived responsiveness and actual resolution of issues.
Another important area for future research is the impact of extrinsic bugs, which propagate through dependency chains, on software ecosystems. Understanding how these bugs affect multiple packages and evaluating the effectiveness of bug reports in addressing them can improve practices for managing such issues. Furthermore, examining how developers and users interpret responsiveness in different contexts can provide insights into enhancing user satisfaction, ensuring that bug resolution strategies effectively balance responsiveness with overall package quality.

\section{Acknowledgment}
This research is partially supported by NSERC 2021~AWD-021280 and JSPS Kakenhi (A) JP24H00692. 

\bibliographystyle{ieeetr}
\bibliography{bibliography}
\end{document}